# Branching ratios of weak hadronic decays of bottom baryons emitting charmless scalar mesons in the pole model

**_Arvind Sharma*_**
_Department of Physics_
_College of Engineering and Management_
_Kapurthala-144601 (INDIA)_

**_Rohit Dhir[†] and R. C. Verma[††]_**
_Department of Physics_
_Punjabi University, Patiala-147002 (INDIA)_
_E-mail: *arvindkamalsharma@gmail.com;_
_[†]dhir.rohit@gmail.com; [††]rcverma@gmail.com._

**ABSTRACT**

We give the first estimate of charmless scalar meson emitting weak hadronic decays of $\Lambda_b^0$, $\Xi_b^0$ and $\Xi_b^-$ bottom baryons employing the pole model and consequently predict their branching ratios.

## I. INTRODUCTION

The heavy baryon mass spectra have become a subject of great interest due to the growing experimental facilities at Belle, BABAR, DELPHI, CLEO, CDF etc. [1-6]. Recently, the lifetime of $\Lambda_b^0$, $\Xi_b^0$ and $\Xi_b^-$ have been measured [7]. Although the experimental data [7] on nonleptonic decays of charm (C=1) baryons have become available in the last decade, measurements on weak decays of bottom baryons have merely begun. On the theoretical side, several authors have investigated weak decays of charm baryons [8-15], only a few attempts have been made [16-19] to study the weak hadronic decays of bottom baryons, mainly emitting *s*-wave mesons. However, the bottom baryons, being heavy, can also emit *p*-wave mesons.

In our recent works [15], we have investigated the *p*-wave meson emitting decays of charmed baryons employing the factorization scheme and including the pole contributions. It has been shown that such decays emitting scalar and axial-vector mesons acquire significant branching ratios of worth observation. In this work, we study the scalar meson emitting decays of bottom baryons. We have already seen that the factorization contribution is negligible in comparison to the pole contributions in case of the scalar meson emitting decays of charmed baryons due to their vanishing decay constants [15c]. For the same reason, factorizable contributions to the bottom baryon decays emitting scalar mesons are also expected to be suppressed. Therefore, we present the first estimate of the branching ratios of weak nonleptonic decays of $\Lambda_b^0$, $\Xi_b^0$ and $\Xi_b^-$ emitting scalar mesons in the pole model.

## II. GENERAL FRAMEWORK

### A. Kinematics

Matrix element for the baryon $B_i(1/2^+) \to B_f(1/2^+) + S_k(0^+)$ decay process can be written as

$$< B_f S_k \mid H_W \mid B_i > = i\bar{u}_{B_f}(A + \gamma_5 B)u_{B_i},$$

where A and B are parity conserving (PC) and parity violating (PV) amplitudes, respectively, $u_B$ are Dirac spinors. Decay width for $B_i(p_i) \to B_f(p_f) + S_k(q)$ is given by

$$\Gamma = C_1[|A|^2 + C_2\,|B|^2], \quad (1)$$

and the asymmetry parameter is

$$\alpha = \frac{2x\,\mathrm{Re}(A^*B)}{|A|^2 + x^2\,|B|^2} \quad (2)$$

where

$$C_1 = \frac{|q_\mu|}{8\pi}\frac{(m_i+m_f)^2 - m_k^2}{m_i^2},$$

$$C_2 = \frac{(m_i-m_f)^2 - m_k^2}{(m_i+m_f)^2 + m_k^2},$$

and $x = q_\mu/(E_f + m_k)$. $E_f$ is the energy of the daughter baryon and four momentum of the scalar meson $q_\mu = (p_i - p_f)_\mu$ is

$$|q_\mu| = \frac{1}{2m_i}\sqrt{[m_i^2 - (m_f - m_k)^2][m_i^2 - (m_f + m_k)^2]},$$

where $m_i$ and $m_f$ are the masses of the initial and final baryons and $m_k$ is the emitted meson mass.

## B. Weak Hamiltonian

For bottom changing $\Delta b = 1$ decays involving $b \to c$ transition, QCD modified current $\otimes$ current weak Hamiltonian is given below:

$$\begin{aligned} H_W = \frac{G_F}{\sqrt{2}}\{ & V_{cb}V_{ud}^*[a_1(\bar{c}b)(\bar{d}u) + a_2(\bar{d}b)(\bar{c}u)] + \\ & V_{cb}V_{cs}^*[a_1(\bar{c}b)(\bar{s}c) + a_2(\bar{s}b)(\bar{c}c)] + \\ & V_{cb}V_{us}^*[a_1(\bar{c}b)(\bar{s}u) + a_2(\bar{s}b)(\bar{c}u)] + \\ & V_{cb}V_{cd}^*[a_1(\bar{c}b)(\bar{d}c) + a_2(\bar{d}b)(\bar{c}c)]\}, \end{aligned} \quad (3)$$

where $(\bar{q}_i q_j) \equiv \bar{q}_i\gamma_\mu(1-\gamma_5)q_j$ denotes the weak *V-A* current. We follow the convention of large $N_c$ limit to fix QCD coefficients $a_1 \approx c_1$ and $a_2 \approx c_2$, where [20]:

$$c_1(\mu) = 1.12 \;,\; c_2(\mu) = -0.26 \text{ at } \mu \approx m_b^2. \quad (4)$$

## C. Scalar meson spectroscopy

The identification of the scalar meson family in the standard nonet picture has been a subject of much controversy. Particle Data Group suggests [7] that there are two sets of scalar mesons nonet, (1) Light scalars: two isoscalars $\sigma(600), f_0(980)$, the isovector $a_0(980)$, and the isodoublet $\kappa(800)$; (2) Heavier scalars: two isoscalars $f_0(1370)$, $f_0(1500)/f_0(1710)$, isovector $a_0(1450)$, and isodoublet $K_0^*(1430)$. In the following, we limit to lighter scalar meson emitting decays of $\Lambda_b^0$, $\Xi_b^0$ and $\Xi_b^-$ .

**(i) $q\bar{q}$ picture**

In the conventional $q\bar{q}$ picture, isovector and isodoublet scalar mesons are given by

$$a_0^+ = u\bar{d}\ ,\ a_0^0 = (u\bar{u} - d\bar{d})/\sqrt{2}\ ,\ a_0^- = d\bar{u}\ ,$$

$$\kappa^+ = u\bar{s}\ ,\ \kappa^0 = d\bar{s}\ ,\ \bar{\kappa}^0 = s\bar{d}\ ,\ \kappa^- = s\bar{u}\ .$$

The unitary singlet and octet states,

$$\varepsilon_1 = (u\bar{u} + d\bar{d} + s\bar{s})/\sqrt{3}\ ,$$

$$\varepsilon_8 = (u\bar{u} + d\bar{d} - 2s\bar{s})/\sqrt{6}\ , \tag{5}$$

mix to generate the physical states as

$$\sigma = \cos\theta_S\varepsilon_1 + \sin\theta_S\varepsilon_8\ ,$$

$$f_0 = -\sin\theta_S\varepsilon_1 + \cos\theta_S\varepsilon_8\ . \tag{6}$$

Alternatively, the mixing can also be expressed as

$$\sigma = (\frac{u\bar{u} + d\bar{d}}{\sqrt{2}})\cos\theta - s\bar{s}\sin\theta\ ,$$

$$f_0 = (\frac{u\bar{u} + d\bar{d}}{\sqrt{2}})\sin\theta + s\bar{s}\cos\theta\ , \tag{7}$$

where $\theta = \pi + (\theta_{ideal} - \theta_S)$. In case of the ideal mixing, $\theta_S = \theta_{ideal} = 35.3^0$ [21], the $s\bar{s}$ component decouples to give

$$\sigma = (u\bar{u} + d\bar{d})/\sqrt{2}\ ,\ f_0 = -s\bar{s}\ ,$$

which is supported by the data of $D_s^+ \to f_0\pi^+$ and $\phi \to f_0\gamma$ implying the copious $f_0(980)$ production via its $s\bar{s}$ component. However, there also exists some experimental evidence indicating that $f_0(980)$ is not purely a $s\bar{s}$ state. $f_0-\sigma$ mixing has been discussed in detail in [22-26], yielding $25^0 < \theta < 40^0$, $140^0 < \theta < 165^0$. In fact, phenomenologically there does not exist a unique mixing angle solution, which may indicate that $\sigma(600)$ and $f_0(980)$ are not purely $q\bar{q}$ bound states.

**(ii) $q^2\bar{q}^2$ picture**

An alternative and arguably more natural explanation for the masses and decay properties of the lightest scalar mesons is to regard these as exotic $q^2\bar{q}^2$ diquark-antidiquark states. In this picture, scalar mesons are given below [21, 26, 27]

$$a_0^+ = u\bar{d}s\bar{s}\,,\ a_0^0 = (sd\bar{s}\bar{d} - su\bar{s}\bar{u})/\sqrt{2}\,,\ a_0^- = d\bar{u}s\bar{s}\,,$$

$$\kappa^+ = ud\bar{s}\bar{d}\,,\ \kappa^0 = ud\bar{s}\bar{u}\,,\ \bar{\kappa}^0 = su\bar{u}\bar{d}\,,\ \kappa^- = sd\bar{u}\bar{d}\,,$$

$$\sigma = ud\bar{u}\bar{d}\,,$$

and

$$f_0 = (sd\bar{s}\bar{d} + su\bar{s}\bar{u})/\sqrt{2}\,.$$

This is supported by a lattice calculation [23] and corresponds to the ideal mixing angle $\tan\theta_S = -\sqrt{2}$ or $\theta_S \approx -54.8^0$ [21]. Similar to the $q\bar{q}$ scenario, general mixing can be described as

$$\sigma = -s\bar{s}(\frac{u\bar{u}+d\bar{d}}{\sqrt{2}})\sin\theta + u\bar{u}d\bar{d}\cos\theta\,,$$

$$f_0 = s\bar{s}(\frac{u\bar{u}+d\bar{d}}{\sqrt{2}})\cos\theta + u\bar{u}d\bar{d}\sin\theta\,, \qquad (8)$$

where $\theta = 174.6^0 \pm 3.3^0$ [28] indicating a small deviation from the ideal mixing angle $(\theta = 180^0)$. However, looking at the uncertainty in determining the angle and for simplicity, we assume ideal mixing in this work.

## III. POLE MODEL

In the pole model, one introduces a set of intermediate states into the decay process so that the weak and strong vertices become separated. In the other way round, the process under consideration passes through certain hadronic intermediate states which can be decomposed into two steps: production of these intermediate states in the strong process, following which the intermediate baryon then undergoes a weak transition to the final baryon. *A* and *B* are then simply given by the product of strong- and weak-coupling constants divided by the mass difference and mass sum, respectively, for *A* and *B*.

For $B_i(1/2^+) \to B_f(1/2^+) + S_k(0^+)$ decay process in *s*- and *u*-channels, positive-parity intermediate baryon $(J^P = 1/2^+)$ poles give rise to the following terms;

$$A^{pole} = -\sum_n \left[ \frac{g_{B_f B_n S_k} a_{ni}}{m_i - m_n} + \frac{a_{fn} g_{B_n B_i S_k}}{m_f - m_n} \right], \tag{9}$$

$$B^{pole} = \sum_n \left[ \frac{g_{B_f B_n S_k} b_{ni}}{m_i + m_n} + \frac{b_{fn} g_{B_n B_i S_k}}{m_f + m_n} \right], \tag{10}$$

where $g_{ijk}$ are the strong baryon-scalar-meson coupling constants. Weak baryon-baryon matrix elements $a_{ij}$ and $b_{ij}$ are defined as

$$< B_i \mid H_W \mid B_j > = \bar{u}_{B_i}(a_{ij} + \gamma_5 b_{ij}) u_{B_j} \,. \tag{11}$$

In addition to the low-lying positive-parity intermediate baryon poles $(J^P = 1/2^+)$, negative-parity intermediate baryon $(J^P = 1/2^-)$ may also contribute to these processes. Unfortunately, there is no information available about the scalar-meson strong coupling constants for the negative-parity baryons. Further, these contributions are expected to be relatively suppressed because of their large masses. Therefore, we have restricted to positive-parity intermediate baryon poles in order to obtain the estimate of the pole contributions to the scalar meson emitting decays of charm baryons. It is well known that the matrix elements $b_{ij}$ vanish for the hyperons in the SU(3) limit [29]. In the case of the charm decays also, it has been shown [8] that $b_{ij} << a_{ij}$, thereby suppressing the PV pole contributions. Assuming the same trend in the bottom sector, PV pole contributions are neglected in the present work. In fact, PV contributions ($B^{pole}$) are further suppressed due to the sum of the baryon masses appearing in the denominator.

**Strong scalar meson-baryon couplings**

In $q\bar{q}$ picture, Hamiltonian representing the strong couplings can be written as

$$H_{strong} = \sqrt{2} g_F (\frac{1}{2} \bar{B}^{[a,b]d} B_{[a,b]c} S_d^c - \bar{B}^{[d,a]b} B_{[a,c]b} S_d^c)$$

$$+\sqrt{2} g_D (\frac{1}{2} \bar{B}^{[a,b]d} B_{[a,b]c} S_d^c + \bar{B}^{[d,a]b} B_{[a,c]b} S_d^c), \quad (12)$$

where $B_{[a,b]c}$ , $\bar{B}^{[a,b]d}$ and $S_d^c$ are the baryon, anti-baryon, and scalar meson tensors respectively and $g_D (g_F)$ are conventional D-type and F-type parameters [30].

On the experimental side, there is no measurement available for the scalar-meson-baryon coupling constants. Recently, G. Erkol *et al.* [21] have obtained the scalar-meson-baryon coupling constants using QCD sum rules. In their analysis, $g_D$ and $g_F$ have been determined as,

$$g_D = 5.4, \text{ and } g_F = 6.6. \quad (13)$$

Similarly, strong couplings ($\bar{B}BS$) have been estimated in $q^2\bar{q}^2$ picture of the scalar mesons in the work [21]. In this case the following values have been obtained

$$g_D = 3.8, \text{ and } g_F = 4.7. \quad (14)$$

The values of strong scalar meson-baryon coupling constants relevant for our calculation have been given in Table I.

**Weak Transitions**

In the tensor notation, the weak Hamiltonian (3) for quark level process $q_i + q_j \to q_l + q_m$ can be expressed as,

$$H_W = \frac{G_F}{\sqrt{2}} V_{il} V_{jm}^* [c_-(m_b) H_{[i,j]}^{[l,m]} + c_+(m_b) H_{(i,j)}^{(l,m)}] \ , \quad (15)$$

where $c_- = c_1 + c_2$ and $c_+ = c_1 - c_2$ and the brackets [,] and (,), respectively, denote the antisymmetrization and symmetrization among the indices. However, for baryon-baryon weak transitions [31], it has been shown that the part of the Hamiltonian $H^{(l,m)}_{(i,j)}$, being symmetric in the color indices also, does not contribute. Thus, by choosing the appropriate indices in the following contraction

$$H_W = a_W [\bar{B}^{[i,j]k} B_{[l,m]k} H^{[l,m]}_{[i,j]}], \tag{16}$$

we obtain the weak baryon-baryon matrix elements ($a_{ij}$) for $\Delta b = 1, \Delta C = 1, \Delta S = 0$ and $\Delta b = 1, \Delta C = 1, \Delta S = -1$ modes which have been given in Table II. It is worth remarking here that since $c$-quark does not appear as constituent in the parent baryons ($\Lambda_b^0$, $\Xi_b^0$, $\Xi_b^-$) considered here, the decays with selection rules $\Delta b = 1, \Delta C = 0, \Delta S = -1$ and $\Delta b = 1, \Delta C = 0, \Delta S = 0$ do not acquire pole contributions from the weak Hamiltonian (3). However, these decay modes may receive contributions through $b + u \to u + s$ and $b + u \to u + d$ quark processes, which are highly suppressed due to the correspondingly small CKM matrix elements.

## IV. NUMERICAL RESULTS: DISCUSSION AND CONCLUSION

We compute the pole contributions using eqn. (9) for $\Delta b = 1, \Delta C = 1, \Delta S = 0$ and $\Delta b = 1, \Delta C = 1, \Delta S = -1$ modes. It may be noted that weak baryon-baryon transitions appearing in pseudoscalar or scalar meson emitting decays of bottom baryons are the same. Sinha *et al.* [17] have already estimated the weak transition amplitudes by quark model calculations as $a_{\Lambda_b^0 \to \Sigma_c^0}$ is related to $a_{\Sigma^+ \to p}$ ( = $1.2 \times 10^{-7}$ GeV) through the following relation:

$$< \Sigma_c^0 | H_W^{PC} | \Lambda_b^0 > = \frac{1}{\sqrt{6}} \frac{V_{cb}}{V_{us}} < p | H_W^{PC} | \Sigma^+ >. \tag{17}$$

However, this estimate is not reliable due to the badly broken SU(5) and ignores the difference in QCD enhancements and flavor dependent baryon overlap function, $|\psi(0)|^2$ appearing in the baryon to baryon weak transitions. Therefore, we follow the quark model analysis of [9, 12, 32], which express

$$< \Sigma_c^0 | H_W^{PC} | \Lambda_b^0 > = 2\sqrt{\frac{2}{3}} \frac{G_F}{\sqrt{2}} c_-(m_b) V_{cb} V_{ud}^* |\psi(0)|_b^2, \tag{18}$$

where $|\psi(0)|_b^2 \equiv <\psi_{\Sigma_c^0}|\delta^3(\vec{r})|\psi_{\Lambda_b^0}>$. Similarly, we obtain

$$<p|H_W^{PC}|\Sigma^+> = -3\frac{G_F}{\sqrt{2}}c_-(m_s)V_{us}V_{ud}^*|\psi(0)|_s^2, \tag{19}$$

where $|\psi(0)|_s^2 \equiv <\psi_p|\delta^3(\vec{r})|\psi_{\Sigma^+}>$. The QCD enhancement due to the hard gluon exchange in the bottom sector $c_-(m_b)$ = 1.38 is lower than that in the charm and hyperon sector with $c_-(m_c)$ = 1.77 and $c_-(m_s) = 2.23$ respectively.

Further, $|\psi(0)|^2$, being a dimensional quantity, may also show variation with flavor [9, 12]. Already, in the study of weak hadronic decays of charmed baryons, this has been estimated through the $\Sigma_c - \Lambda_c$ hyperfine splitting. Similarly, using the constituent quark model [9, 12, 33], the following ratio of the hyperfine splitting in the strange and bottom sectors

$$\frac{\Sigma_b - \Lambda_b}{\Sigma - \Lambda} = \frac{\alpha_s(m_b)}{\alpha_s(m_s)}\frac{m_s(m_b - m_u)|\psi(0)|_b^2}{m_b(m_s - m_u)|\psi(0)|_s^2}, \tag{20}$$

yields

$$R \equiv \frac{|\psi(0)|_b^2}{|\psi(0)|_s^2} \approx 2.27, \tag{21}$$

for the choice $\alpha_s(m_b)/\alpha_s(m_s) \approx 0.40$. Finally branching ratios are evaluated without and with $|\psi(0)|^2$ variation, which are presented in Tables III-V and Tables VI-VIII in both $q\bar{q}$ and $q^2\bar{q}^2$ pictures respectively. We observe the following:

(1) In both the $q\bar{q}$ and $q^2\bar{q}^2$ pictures, the dominant decays modes are $\Lambda_b^0 \to \Sigma_c^+ a_0^- / \Sigma_c^0 a_0^0 / \Sigma_c^0 \sigma / \Xi_c^{'0} \kappa_0^0$, $\Xi_b^0 \to \Xi_c^{'+} a_0^- / \Xi_c^{'0} \sigma / \Omega_c^0 \kappa_0^0$ and $\Xi_b^0 \to \Sigma_c^0 \kappa_0^-$ with branching ratios of the order of $10^{-3} - 10^{-4}$, hopefully within the reach of experimental observation.

(2) However, the decay $\Lambda_b^0 \to \Sigma_c^0 f_0$ forbidden in $q\bar{q}$ picture of scalar mesons acquire a non-zero branching ratio around $2.89\times10^{-3}$ in $q^2\bar{q}^2$ picture. This provides a useful test for the 4-quark picture of the scalar mesons.

(3) All decays of $\Delta b = 1, \Delta C = 1, \Delta S = -1$ mode are suppressed in comparison to $\Delta b = 1, \Delta C = 1, \Delta S = 0$ mode due to the small value of CKM matrix elements

(4) Asymmetry parameters for all decays vanish due to suppressed weak PV transition amplitudes $b_{ij}$'s.

(5) Branching ratios of all the decays, in both the pictures, get enhanced by a factor of five due to the possible flavor dependence of $|\psi(0)|^2$ appearing in the baryon-baryon weak transition amplitudes.

(6) It is also noted that the decays with selection rules $\Delta b = 1, \Delta C = 0, \Delta S = -1$ and $\Delta b = 1, \Delta C = 0, \Delta S = 0$ do not acquire pole contributions as *c*-quark does not appear as constituent in the parent baryons ($\Lambda_b^0$, $\Xi_b^0$, $\Xi_b^-$) considered in this work. However, these decay modes may receive contributions through $b + u \to u + s$ and $b + u \to u + d$ quark processes, which are highly suppressed due to the correspondingly small CKM matrix elements.

Table I: Scalar-meson-baryon strong coupling constants

| $B \to BS$ | $q\bar{q}$ picture | $q^2\bar{q}^2$ picture |
|---|---|---|
| $\Xi_c^0 \to \Lambda_c^+ \kappa_0^-$ | -4.2 | -3.1 |
| $\Xi_c^0 \to \Sigma_c^o \kappa_0^0$ | 6.2 | 4.4 |
| $\Xi_c^0 \to \Xi_c^0 a_0^0$ | -3.0 | 2.2 |
| $\Xi_c^0 \to \Xi_c^0 \sigma$ | -4.3 | 3.1 |
| $\Xi_c^0 \to \Xi_c^0 f_0$ | -7.0 | 6.5 |
| $\Xi_c^0 \to \Xi_c^{'0} a_0^0$ | 3.1 | -2.2 |
| $\Xi_c^0 \to \Xi_c^{'0} \sigma$ | 4.4 | -3.1 |
| $\Xi_c^0 \to \Xi_c^{'0} f_0$ | -3.1 | 2.2 |
| $\Xi_c^0 \to \Omega_c^0 \kappa_0^0$ | -6.2 | -4.4 |
| | | |
| $\Xi_c^{'0} \to \Lambda_c^+ \kappa_0^-$ | -4.4 | -3.1 |
| $\Xi_c^{'0} \to \Sigma_c^0 \kappa_0^0$ | 13.2 | 9.4 |
| $\Xi_c^{'0} \to \Xi_c^{'0} a_0^0$ | -6.6 | 4.7 |
| $\Xi_c^{'0} \to \Xi_c^{'0} \sigma$ | 6.6 | 6.6 |
| $\Xi_c^{'0} \to \Xi_c^{'0} f_0$ | -9.3 | 14.1 |
| $\Xi_c^{'0} \to \Omega_c^0 \kappa_0^0$ | 13.2 | 9.4 |
| | | |
| $\Sigma_c^0 \to \Lambda_c^+ a_0^-$ | -6.2 | -4.4 |
| $\Sigma_c^0 \to \Sigma_c^0 a_0^0$ | -13.2 | 9.4 |
| $\Sigma_c^0 \to \Sigma_c^0 \sigma$ | 13.2 | 13.3 |
| $\Sigma_c^0 \to \Sigma_c^0 f_0$ | -13.2 | 9.4 |
| $\Sigma_c^0 \to \Xi_c^0 \kappa_0^0$ | 6.2 | 4.4 |
| | | |
| $\Omega_c^0 \to \Xi_c^+ \kappa_0^-$ | -6.2 | -4.4 |
| $\Omega_c^0 \to \Xi_c^{'+} \kappa_0^-$ | 13.2 | 9.4 |
| $\Omega_c^0 \to \Omega_c^0 a_0^0$ | 0 | 0 |
| $\Omega_c^0 \to \Omega_c^0 \sigma$ | 0 | 0 |
| $\Omega_c^0 \to \Omega_c^0 f_0$ | 18.6 | 18.8 |
| | | |
| $\Lambda_b^0 \to \Lambda_b^0 a_0^0$ | 0 | 0 |
| $\Lambda_b^0 \to \Lambda_b^0 \sigma$ | 6.0 | 6.1 |
| $\Lambda_b^0 \to \Lambda_b^0 f_0$ | 0 | 4.3 |
| $\Lambda_b^0 \to \Sigma_b^0 a_0^0$ | -6.2 | 4.4 |
| $\Lambda_b^0 \to \Sigma_b^0 \sigma$ | 0 | 0 |
| $\Lambda_b^0 \to \Sigma_b^0 f_0^0$ | 0 | 0 |
| $\Lambda_b^0 \to \Xi_b^0 \kappa_0^0$ | 4.2 | 3.0 |
| $\Lambda_b^0 \to \Xi_b^{'0} \kappa_0^-$ | 4.4 | 3.1 |
| | | |
| $\Xi_b^0 \to \Sigma_b^0 \kappa_0^0$ | 4.4 | 3.1 |
| $\Xi_b^0 \to \Xi_b^0 a_0^0$ | 3.0 | -2.2 |
| $\Xi_b^0 \to \Xi_b^0 \sigma$ | 3.0 | 3.1 |
| $\Xi_b^0 \to \Xi_b^0 f_0$ | -4.2 | 6.5 |
| $\Xi_b^0 \to \Xi_b^{'0} a_0^0$ | -3.1 | 2.2 |
| $\Xi_b^0 \to \Xi_b^{'0} \sigma$ | 4.4 | -3.1 |
| $\Xi_b^0 \to \Xi_b^{'0} f_0^0$ | -3.1 | 2.2 |
| | | |
| $\Xi_b^- \to \Lambda_b^0 \kappa_0^-$ | -4.2 | -3.0 |
| $\Xi_b^- \to \Sigma_b^0 \kappa_0^-$ | 4.4 | 3.1 |
| $\Xi_b^- \to \Xi_b^0 a_0^-$ | 4.2 | 3.1 |
| $\Xi_b^- \to \Xi_b^{'0} a_0^-$ | -4.4 | -3.1 |

Table II: Weak baryon-baryon transition amplitudes

| Weak transition | Transition amplitude ($\times a_W$) |
|---|---|
| $\Delta b = 1, \Delta C = 1, \Delta S = 0$ | |
| $\Lambda_b^0 \to \Sigma_c^0$ | $\sqrt{3/2}$ |
| $\Sigma_b^+ \to \Lambda_c^+$ | $-\sqrt{3/2}$ |
| $\Sigma_b^+ \to \Sigma_c^+$ | $3/\sqrt{2}$ |
| $\Sigma_b^0 \to \Sigma_c^0$ | $3/\sqrt{2}$ |
| $\Xi_b^0 \to \Xi_c^0$ | $1/2$ |
| $\Xi_b^0 \to \Xi_c^{'0}$ | $\sqrt{3}/2$ |
| $\Xi_b^{'0} \to \Xi_c^0$ | $\sqrt{3}/2$ |
| $\Xi_b^{'0} \to \Xi_c^{'0}$ | $3/2$ |
| $\Delta b = 1, \Delta C = 1, \Delta S = -1$ | |
| $\Lambda_b^0 \to \Xi_c^0$ | $-1/2$ |
| $\Lambda_b^0 \to \Xi_c^{'0}$ | $\sqrt{3}/2$ |
| $\Sigma_b^+ \to \Xi_c^+$ | $-\sqrt{3/2}$ |
| $\Sigma_b^+ \to \Xi_c^{'+}$ | $3/\sqrt{2}$ |
| $\Sigma_b^0 \to \Xi_c^0$ | $-\sqrt{3}/2$ |
| $\Sigma_b^0 \to \Xi_c^{'0}$ | $3/2$ |
| $\Xi_b^0 \to \Omega_c^0$ | $\sqrt{3/2}$ |
| $\Xi_b^{'0} \to \Omega_c^0$ | $3/\sqrt{2}$ |

Table III: Branching ratio for $\Lambda_b^0$ decays ($q\bar{q}$ picture)

| Decay | Branching Ratio (%) without $\vert\psi(0)\vert^2$ variation | Branching Ratio (%) with $\vert\psi(0)\vert^2$ variation |
|---|---|---|
| $\Delta b=1, \Delta C=1, \Delta S=0$ | | |
| $\Lambda_b^0 \to \Lambda_c^+ a_0^-$ | $4.52\times10^{-5}$ | $2.34\times10^{-4}$ |
| $\Lambda_b^0 \to \Xi_c^0 \kappa_0^0$ | $3.50\times10^{-4}$ | $1.81\times10^{-3}$ |
| $\Lambda_b^0 \to \Sigma_c^+ a_0^-$ | $1.03\times10^{-3}$ | $5.33\times10^{-3}$ |
| $\Lambda_b^0 \to \Sigma_c^0 a_0^0$ | $1.03\times10^{-3}$ | $5.32\times10^{-3}$ |
| $\Lambda_b^0 \to \Sigma_c^0 \sigma$ | $6.13\times10^{-3}$ | $3.17\times10^{-2}$ |
| $\Lambda_b^0 \to \Xi_c^{'0} \kappa_0^0$ | $3.08\times10^{-3}$ | $1.59\times10^{-2}$ |
| $\Delta b=1, \Delta C=1, \Delta S=-1$ | | |
| $\Lambda_b^0 \to \Lambda_c^+ \kappa_0^-$ | $1.34\times10^{-5}$ | $6.96\times10^{-5}$ |
| $\Lambda_b^0 \to \Xi_c^+ a_0^-$ | $4.91\times10^{-6}$ | $2.53\times10^{-5}$ |
| $\Lambda_b^0 \to \Xi_c^0 a_0^0$ | $2.48\times10^{-6}$ | $1.28\times10^{-5}$ |
| $\Lambda_b^0 \to \Xi_c^0 \sigma$ | $6.97\times10^{-6}$ | $3.60\times10^{-5}$ |
| $\Lambda_b^0 \to \Xi_c^0 f_0$ | $1.33\times10^{-5}$ | $6.88\times10^{-5}$ |
| $\Lambda_b^0 \to \Sigma_c^+ \kappa_0^-$ | $1.53\times10^{-4}$ | $7.94\times10^{-4}$ |
| $\Lambda_b^0 \to \Sigma_c^0 \bar{\kappa}_0^0$ | $3.07\times10^{-4}$ | $1.58\times10^{-3}$ |
| $\Lambda_b^0 \to \Xi_c^{'+} a_0^-$ | $2.08\times10^{-5}$ | $1.07\times10^{-4}$ |
| $\Lambda_b^0 \to \Xi_c^{'0} a_0^0$ | $1.05\times10^{-5}$ | $5.43\times10^{-5}$ |
| $\Lambda_b^0 \to \Xi_c^{'0} \sigma$ | $1.83\times10^{-5}$ | $9.48\times10^{-5}$ |
| $\Lambda_b^0 \to \Xi_c^{'0} f_0$ | $1.51\times10^{-4}$ | $7.80\times10^{-4}$ |
| $\Xi_b^0 \to \Omega_c^0 \kappa_0^0$ | $1.48\times10^{-6}$ | $7.69\times10^{-6}$ |

**Table IV: Branching ratio for $\Xi_b^0$ decays ($q\bar{q}$ picture)**

| Decay | Branching Ratio (%) without $\|\psi(0)\|^2$ variation | Branching Ratio (%) with $\|\psi(0)\|^2$ variation |
|---|---|---|
| $\Delta b = 1, \Delta C = 1, \Delta S = 0$ | | |
| $\Xi_b^0 \to \Lambda_c^+ \kappa_0^-$ | $9.00\times10^{-5}$ | $4.64\times10^{-4}$ |
| $\Xi_b^0 \to \Xi_c^+ a_0^-$ | $2.35\times10^{-4}$ | $1.21\times10^{-3}$ |
| $\Xi_b^0 \to \Xi_c^0 a_0^0$ | $3.96\times10^{-4}$ | $2.05\times10^{-3}$ |
| $\Xi_b^0 \to \Xi_c^0 \sigma$ | $3.30\times10^{-6}$ | $1.71\times10^{-5}$ |
| $\Xi_b^0 \to \Xi_c^0 f_0$ | $6.34\times10^{-6}$ | $3.27\times10^{-5}$ |
| $\Xi_b^0 \to \Sigma_c^+ \kappa_0^-$ | $4.51\times10^{-4}$ | $2.34\times10^{-3}$ |
| $\Xi_b^0 \to \Sigma_c^0 \bar{\kappa}_0^0$ | $4.19\times10^{-6}$ | $2.16\times10^{-5}$ |
| $\Xi_b^0 \to \Xi_c^{'+} a_0^-$ | $2.68\times10^{-3}$ | $1.39\times10^{-2}$ |
| $\Xi_b^0 \to \Xi_c^{'0} a_0^0$ | $4.17\times10^{-4}$ | $2.15\times10^{-3}$ |
| $\Xi_b^0 \to \Xi_c^{'0} \sigma$ | $2.91\times10^{-3}$ | $1.51\times10^{-2}$ |
| $\Xi_b^0 \to \Xi_c^{'0} f_0$ | $4.23\times10^{-6}$ | $2.19\times10^{-5}$ |
| $\Xi_b^0 \to \Omega_c^0 \kappa_0^0$ | $5.47\times10^{-3}$ | $2.83\times10^{-2}$ |
| $\Delta b = 1, \Delta C = 1, \Delta S = -1$ | | |
| $\Xi_b^0 \to \Xi_c^+ \kappa_0^-$ | $1.42\times10^{-6}$ | $7.34\times10^{-6}$ |
| $\Xi_b^0 \to \Xi_c^0 \bar{\kappa}_0^0$ | $1.75\times10^{-5}$ | $9.09\times10^{-5}$ |
| $\Xi_b^0 \to \Xi_c^{'+} \kappa_0^-$ | $5.39\times10^{-5}$ | $2.78\times10^{-4}$ |
| $\Xi_b^0 \to \Xi_c^{'0} \bar{\kappa}_0^0$ | $1.64\times10^{-4}$ | $8.35\times10^{-4}$ |
| $\Xi_b^0 \to \Omega_c^0 a_0^0$ | $2.92\times10^{-5}$ | $1.51\times10^{-4}$ |
| $\Xi_b^0 \to \Omega_c^0 \sigma$ | $3.07\times10^{-5}$ | $1.59\times10^{-4}$ |
| $\Xi_b^0 \to \Omega_c^0 f_0$ | $3.27\times10^{-4}$ | $1.69\times10^{-3}$ |

**Table V: Branching ratio for $\Xi_b^-$ decays ($q\bar{q}$ picture)**

| Decay | Branching Ratio (%) without $\lvert\psi(0)\rvert^2$ variation | Branching Ratio (%) with $\lvert\psi(0)\rvert^2$ variation |
|---|---|---|
| $\Delta b = 1, \Delta C = 1, \Delta S = 0$ | | |
| $\Xi_b^- \to \Xi_c^0 a_0^-$ | $1.64\times10^{-4}$ | $8.50\times10^{-4}$ |
| $\Xi_b^- \to \Sigma_c^0 \kappa_0^-$ | $1.03\times10^{-3}$ | $5.35\times10^{-3}$ |
| $\Xi_b^- \to \Xi_c^{'0} a_0^-$ | $5.24\times10^{-4}$ | $2.71\times10^{-3}$ |
| $\Delta b = 1, \Delta C = 1, \Delta S = -1$ | | |
| $\Xi_b^- \to \Xi_c^0 \kappa_0^-$ | $9.08\times10^{-6}$ | $4.69\times10^{-5}$ |
| $\Xi_b^- \to \Xi_c^{'0} \kappa_0^-$ | $2.90\times10^{-5}$ | $1.50\times10^{-4}$ |
| $\Xi_b^- \to \Omega_c^0 a_0^-$ | $5.84\times10^{-5}$ | $3.02\times10^{-4}$ |

**Table VI: Branching ratio for $\Lambda_b^0$ decays ($q^2\bar{q}^2$ picture)**

| Decay | Branching Ratio (%) without $\lvert\psi(0)\rvert^2$ variation | Branching Ratio (%) with $\lvert\psi(0)\rvert^2$ variation |
|---|---|---|
| $\Delta b = 1, \Delta C = 1, \Delta S = 0$ | | |
| $\Lambda_b^0 \to \Lambda_c^+ a_0^-$ | $2.24\times10^{-5}$ | $1.15\times10^{-4}$ |
| $\Lambda_b^0 \to \Xi_c^0 \kappa_0^0$ | $1.64\times10^{-4}$ | $8.52\times10^{-4}$ |
| $\Lambda_b^0 \to \Sigma_c^+ a_0^-$ | $5.66\times10^{-4}$ | $2.92\times10^{-3}$ |
| $\Lambda_b^0 \to \Sigma_c^0 a_0^0$ | $5.65\times10^{-4}$ | $2.92\times10^{-3}$ |
| $\Lambda_b^0 \to \Sigma_c^0 \sigma$ | $6.06\times10^{-3}$ | $3.14\times10^{-2}$ |
| $\Lambda_b^0 \to \Sigma_c^0 f_0$ | $2.89\times10^{-3}$ | $1.49\times10^{-2}$ |
| $\Lambda_b^0 \to \Xi_c^{'0} \kappa_0^0$ | $1.57\times10^{-3}$ | $8.14\times10^{-3}$ |
| $\Delta b = 1, \Delta C = 1, \Delta S = -1$ | | |
| $\Lambda_b^0 \to \Lambda_c^+ \kappa_0^-$ | $6.26\times10^{-6}$ | $3.23\times10^{-5}$ |
| $\Lambda_b^0 \to \Xi_c^+ a_0^-$ | $2.19\times10^{-6}$ | $1.13\times10^{-5}$ |
| $\Lambda_b^0 \to \Xi_c^0 a_0^0$ | $1.11\times10^{-6}$ | $5.74\times10^{-6}$ |
| $\Lambda_b^0 \to \Xi_c^0 \sigma$ | $6.49\times10^{-6}$ | $3.35\times10^{-5}$ |
| $\Lambda_b^0 \to \Xi_c^0 f_0$ | $3.09\times10^{-6}$ | $1.60\times10^{-5}$ |
| $\Lambda_b^0 \to \Sigma_c^+ \kappa_0^-$ | $7.85\times10^{-5}$ | $4.06\times10^{-4}$ |
| $\Lambda_b^0 \to \Sigma_c^0 \bar{\kappa}_0^0$ | $1.57\times10^{-4}$ | $8.12\times10^{-4}$ |
| $\Lambda_b^0 \to \Xi_c^{'+} a_0^-$ | $9.44\times10^{-6}$ | $4.88\times10^{-5}$ |
| $\Lambda_b^0 \to \Xi_c^{'0} a_0^0$ | $4.76\times10^{-6}$ | $2.46\times10^{-5}$ |
| $\Lambda_b^0 \to \Xi_c^{'0} \sigma$ | $1.69\times10^{-5}$ | $8.76\times10^{-5}$ |
| $\Lambda_b^0 \to \Xi_c^{'0} f_0$ | $2.33\times10^{-4}$ | $1.20\times10^{-3}$ |
| $\Xi_b^0 \to \Omega_c^0 \kappa_0^0$ | $7.42\times10^{-7}$ | $3.84\times10^{-6}$ |

**Table VII: Branching ratio for $\Xi_b^0$ decays ($q^2\bar{q}^2$ picture)**

| Decay | Branching Ratio (%) without $\lvert\psi(0)\rvert^2$ variation | Branching Ratio (%) with $\lvert\psi(0)\rvert^2$ variation |
|---|---|---|
| $\Delta b = 1, \Delta C = 1, \Delta S = 0$ | | |
| $\Xi_b^0 \to \Lambda_c^+ \kappa_0^-$ | $4.01\times10^{-5}$ | $2.07\times10^{-4}$ |
| $\Xi_b^0 \to \Xi_c^+ a_0^-$ | $1.09\times10^{-4}$ | $5.65\times10^{-4}$ |
| $\Xi_b^0 \to \Xi_c^0 a_0^0$ | $1.83\times10^{-4}$ | $9.48\times10^{-4}$ |
| $\Xi_b^0 \to \Xi_c^0 \sigma$ | $3.27\times10^{-6}$ | $1.69\times10^{-5}$ |
| $\Xi_b^0 \to \Xi_c^0 f_0$ | $1.57\times10^{-6}$ | $8.11\times10^{-6}$ |
| $\Xi_b^0 \to \Sigma_c^+ \kappa_0^-$ | $2.06\times10^{-4}$ | $1.06\times10^{-3}$ |
| $\Xi_b^0 \to \Sigma_c^0 \bar{\kappa}_0^0$ | $2.06\times10^{-6}$ | $1.06\times10^{-5}$ |
| $\Xi_b^0 \to \Xi_c^{'+} a_0^-$ | $1.37\times10^{-3}$ | $7.08\times10^{-3}$ |
| $\Xi_b^0 \to \Xi_c^{'0} a_0^0$ | $2.31\times10^{-4}$ | $1.19\times10^{-3}$ |
| $\Xi_b^0 \to \Xi_c^{'0} \sigma$ | $2.88\times10^{-3}$ | $1.49\times10^{-2}$ |
| $\Xi_b^0 \to \Xi_c^{'0} f_0$ | $1.53\times10^{-3}$ | $7.95\times10^{-3}$ |
| $\Xi_b^0 \to \Omega_c^0 \kappa_0^0$ | $2.80\times10^{-3}$ | $1.44\times10^{-2}$ |
| $\Delta b = 1, \Delta C = 1, \Delta S = -1$ | | |
| $\Xi_b^0 \to \Xi_c^+ \kappa_0^-$ | $7.03\times10^{-7}$ | $3.64\times10^{-6}$ |
| $\Xi_b^0 \to \Xi_c^0 \bar{\kappa}_0^0$ | $8.24\times10^{-6}$ | $4.26\times10^{-5}$ |
| $\Xi_b^0 \to \Xi_c^{'+} \kappa_0^-$ | $2.97\times10^{-5}$ | $1.53\times10^{-4}$ |
| $\Xi_b^0 \to \Xi_c^{'0} \bar{\kappa}_0^0$ | $8.25\times10^{-5}$ | $4.26\times10^{-4}$ |
| $\Xi_b^0 \to \Omega_c^0 a_0^0$ | $1.34\times10^{-5}$ | $6.95\times10^{-5}$ |
| $\Xi_b^0 \to \Omega_c^0 \sigma$ | $2.82\times10^{-5}$ | $1.46\times10^{-4}$ |
| $\Xi_b^0 \to \Omega_c^0 f_0$ | $4.82\times10^{-4}$ | $2.49\times10^{-3}$ |

**Table VIII: Branching ratio for $\Xi_b^-$ decays ($q^2\bar{q}^2$ picture)**

| Decay | Branching Ratio (%) without $\lvert\psi(0)\rvert^2$ variation | Branching Ratio (%) with $\lvert\psi(0)\rvert^2$ variation |
|---|---|---|
| $\Delta b = 1, \Delta C = 1, \Delta S = 0$ | | |
| $\Xi_b^- \to \Xi_c^0 a_0^-$ | $7.55\times10^{-5}$ | $3.90\times10^{-4}$ |
| $\Xi_b^- \to \Sigma_c^0 \kappa_0^-$ | $4.74\times10^{-4}$ | $2.45\times10^{-3}$ |
| $\Xi_b^- \to \Xi_c^{'0} a_0^-$ | $2.40\times10^{-4}$ | $1.24\times10^{-3}$ |
| $\Delta b = 1, \Delta C = 1, \Delta S = -1$ | | |
| $\Xi_b^- \to \Xi_c^0 \kappa_0^-$ | $4.16\times10^{-6}$ | $2.15\times10^{-5}$ |
| $\Xi_b^- \to \Xi_c^{'0} \kappa_0^-$ | $1.33\times10^{-5}$ | $6.88\times10^{-5}$ |
| $\Xi_b^- \to \Omega_c^0 a_0^-$ | $2.69\times10^{-5}$ | $1.39\times10^{-4}$ |

## REFERENCES

1. R. Mizuk *et al.* (Belle Collaboration) Phys. Rev. Lett. **94**, 122002 (2005).
2. B. Aubert *et al.* (BABAR Collaboration) Phys. Rev. Lett. **98**, 122011 (2007).
3. M. Feindt *et al.* (DELPHI Colaboration) Report no. CERN-PRE/95-139 (2007).
4. K. W. Edwards *et al.* (CLEO Collaboration) Phys. Rev. Lett. **74**, 3331 (1995).
5. M. Artuso *et al.* (CLEO Collaboration) Phys. Rev. Lett. **86**, 4479 (2001).
6. I. V. Gorelov (CDF Collaboration) arXiv:hep-ex/0701056,(2007).
7. C. Amsler *et al.* (Particle Data Group), Phys. Lett. B**667**, 1 (2008).
8. H. Y. Cheng and B. Tseng, Phys. Rev. D **46**, 1042 (1992).
9. T. Uppal, R. C. Verma and M. P. Khanna, Phys. Rev. D **49**, 3417 (1994).
10. M. A. Ivanov *et al.*, Phys. Rev. D **57**, 5632 (1998).
11. B. Guberina and B. Melic, Eur. Phys. J. C **2**, 697 (1998); B. Guberina and H. Stefancic, Phys. Rev. D **65**, 114004 (2002).
12. K. K. Sharma and R. C. Verma, Eur. J. C **7**, 217 (1999).
13. R.C.Verma, Ind. J. Pure Appl. Phys. **38**, 395 (2000); E. N. Bukina *et al.*, Nuclear Phys. **B93**, 34 (2001).
14. V. E. Lyubovitskij *et al.*, Prog. Part. Nucl. Phys. **50**, 329 (2003).
15. A. Sharma and R. C. Verma, Phys. Rev. D **71**, 074024 (2005); ibid **79**, 037506 (2009); A. Sharma and R. C. Verma, J. Phys. G: Nucl. Part. Phys. **36**, 075005 (2009).
16. H. Y. Cheng, Phys. Rev. D **56**, 2799 (1997).
17. Sonali Sinha, M. P. Khanna and R. C. Verma, Phys. Rev. D **57**, 4483 (1998).
18. Fayyazuddin and Riazuddin, Phys. Rev. D **58**, 014016 (1998).
19. M. R. Khodja, Fayyazuddin and Riazuddin, Phys. Rev. D **60**, 053005 (1999).
20. A. Ali and C. Greub, Phys. Rev. D **57**, 2996 (1998).
21. G. Erkol *et al.*, Phys. Rev. C **73**, 044009 (2006).
22. H. Y. Cheng, Phys. Rev. D **67**, 034024 (2003).
23. M. Alford and R. L. Jaffe, Nucl. Phys. **B578**, 367 (2000).

24. A. V. Anisovish *et al.*, Eur. Phys. J A **12**, 103 (2001); Phys. At. Nucl. **65**, 497 (2002).
25. A. Gokalp *et al.,* Phys. Lett. **609B**, 291 (2005).
26. H. Y. Cheng *et al.*, Phys. Rev. D **73**, 014017 (2006).
27. R. L. Jaffe, Phys. Rev. D **15**, 267 (1977); ibid. **15**, 281 (1977); ibid. **17**, 1444 (1978).
28. L. Maini *et al.*, Phys. Rev. Lett. **93**, 212002 (2004).
29. R. E. Marshak, X. Riazzudin and C. P. Ryan, *Theory of Weak Interactions in Particle Physics* (Wiley, New York, 1969); M. D. Scadron and L. R. Thebaud, Phys. Rev. D **8**, 2190 (1973); H. Y. Cheng, Phys. Rev. D **56**, 2799 (1997); A. Babic *et al.*, Phys. Rev. D **70**, 117501 (2004).Riazuddin and Fayyazuddin, Phys. Rev. D **18**, 1578 (1978); ibid. **55**, 255 (1997).
30. J. G. Korner *et al.*, Z. Phys. C **1**, 69 (1979); J. G. Korner and M. Krammer, ibid. **55**, 659 (1992); M. P. Khanna, Phys. Rev. D **49**, 5921 (1994).
31. J. G. Korner *et al.,* Phys. Lett**. 78B**, 492 (1978); B. Guberina *et al.,* Z. Phys. C **13**, 251 (1982); H. Y. Cheng, Z. Phys. C **29**, 453 (1985); J. Donoghue, Phys. Rev. D **33**, 1516 (1986); M. J. Savage and R. P. Springer, Phys. Rev. D **42**, 1527 (1990); S. Pakvasa *et al.*, ibid. **42**, 3746 (1990); R. E. Karlsen and M. D. Scadron, Europhys. Lett. **14**, 319 (1991).
32. Riazuddin and Fayyazuddin, Phys. Rev. D **18**, 1578 (1978); ibid. **55**, 255 (1997).
33. A. de Rujula, H. Georgi and S. Glashow, Phys. Rev. D **12**, 147 (1975); D. H. Perkins, *Introduction to High Energy Physics* (Cambridge University Press, Cambridge, 2000).